\documentclass[twocolumn,floatfix]{revtex4}
\pdfoutput=1
\usepackage{graphicx}
\usepackage{dcolumn}
\usepackage{longtable}
\usepackage{amsmath}
\usepackage{amssymb}

\begin{document}
\title{Shell model structure of proxy-SU(3) pairs of orbitals
}

\author
{Dennis Bonatsos$^1$, Hadi Sobhani$^2$, and Hassan Hassanabadi$^2$  }

\affiliation
{$^1$Institute of Nuclear and Particle Physics, National Centre for Scientific Research 
``Demokritos'', GR-15310 Aghia Paraskevi, Attiki, Greece}

\affiliation
{$^2$ Faculty of Physics, Shahrood University of Technology, Shahrood, Iran, P.O. Box 3619995161-316}

\begin{abstract}

The Nilsson orbitals used in the substitutions occurring in the proxy-SU(3) scheme, which are the orbitals bearing the maximum value of total angular momentum in each shell, have an extremely simple structure in the shell model basis $|N l j \Omega \rangle$, with each Nilsson orbital corresponding to a single shell model eigenvector. This simple structure is valid at all deformations for these orbitals, while in other orbitals it is valid only at small deformations. Nilsson 0[110] pairs are found to correspond to $|1 1 1 0\rangle$ pairs in the spherical shell model basis, paving the way for using the proxy-SU(3) approximation within the shell model.  

 \end{abstract}

\maketitle
      
\section{Introduction} 

SU(3) symmetry has been playing an important role in nuclear physics for a long time \cite{Kota}. After its discovery by Elliott \cite{Elliott1,Elliott2,Elliott3} in the nuclear sd shell, it has been used in the framework of several algebraic models using bosons, like the Interacting Boson Model \cite{IA,IVI,FVI} and the Vector Boson Model \cite{Afanasev,Minkov}, or fermions, like the symplectic model \cite{Rosensteel,RW} and the Fermion Dynamic Symmetry Model \cite{FDSM}. Furthermore, approximate SU(3) symmetries have been introduced in heavier shells, including the pseudo-SU(3) \cite{pseudo1,pseudo2}, quasi-SU(3) \cite{Zuker1,Zuker2}, and proxy-SU(3) \cite{proxy1,proxy2} schemes. SU(3) techniques are also used within large-scale no-core shell model calculations \cite{Launey1,Launey2} in order to reduce their size. 

In the present work we are going to focus attention on the recently introduced proxy-SU(3) scheme \cite{proxy1}, which has provided successful parameter-free predictions \cite{proxy2} for the collective variables $\beta$ and $\gamma$, expressing the deviation of atomic nuclei from the spherical shape and from axial symmetry respectively \cite{BM}. In addition, proxy-SU(3) has provided \cite{proxy2,proxy3} a solution to the long-standing puzzle \cite{Hamamoto} of the dominance of prolate over oblate shapes in the ground state bands of even-even nuclei, as well as a prediction for the prolate to oblate shape phase transition in the heavy rare earths, in agreement to experimental evidence \cite{Linnemann}. Shell-like quarteting in heavy nuclei \cite{Cseh} has also been considered recently within the proxy-SU(3) approach.  

Nuclear shells are known to be derived in the nuclear shell model \cite{Mayer1,MJ} from three-dimensional harmonic oscillator (3D-HO) shells bearing U(N) symmetries possessing SU(3) subalgebras \cite{Wybourne,Smirnov,IacLie,BK}. These symmetries are broken beyond the sd shell by the strong spin-orbit interaction, which influences maximally within each shell the orbitals bearing the highest total angular momentum $j$, which are pushed by the spin-orbit interaction  down into the shell below, replaced by the relevant orbitals coming from the shell above.    
The basic idea behind proxy-SU(3) is the restoration of the U(N) symmetry of the 3D-HO by replacing in each shell the intruder orbitals, invading from the shell above, by the orbitals who have deserted this shell by going into the shell below, which are therefore playing the role of proxies of the intruder orbitals.  

The proxy-SU(3) scheme has been introduced \cite{proxy1} in the framework of the Nilsson model \cite{Nilsson1,NR}, which, despite its simplicity, provides successful predictions for the single-particle spectra of deformed atomic nuclei, which have been extremely useful over many years for the interpretation of experimental results. Single particle orbitals in the Nilsson model are labeled by  $\Omega [N n_z \Lambda]$,  where $N$ is the principal quantum number of the major shell,
$n_z$ is the number of nodes in the wave function in the $z$-direction, while $\Lambda$ and $\Omega$ are the projections 
of the orbital angular momentum and the total angular momentum respectively on the $z$-axis. Proxy-SU(3) is based on the replacement of the intruder orbitals (except the one with the highest projection $K$ of the total angular momentum $j$) by the orbitals which have deserted this shell by going into the shell below, from which they differ by $\Delta \Omega [\Delta N \Delta n_z \Delta \Lambda]= 0[110]$. We call these orbitals 0[110] pairs. 

0[110] proton-neutron pairs have first been singled out experimentally in \cite{Cakirli} as the pairs corresponding to maximum proton-neutron interaction, as determined from double differences of binding energies. It was later shown that they correspond to wave functions exhibiting large spatial overlaps \cite{Karampagia}. All orbitals within each valence shell were taken into account in these studies (see Fig. 4 of \cite{Karampagia}), demonstrating  the 0[110] correspondence between the proton orbitals and their neutron partners.

In proxy-SU(3), 0[110] pairs are used, but there are three main differences: a) pairs of protons alone or neutrons alone are used, b) not all valence nucleons are involved in the approximation (replacement) made, but only the intruder orbitals within each valence shell do, c) within each valence shell the intruder orbitals are replaced by the orbitals which have escaped into the shell below, while no replacement is made in \cite{Karampagia}. 

It has become clear that proxy-SU(3) predictions \cite{proxy2} appear to be correct outside the region of deformed nuclei, in which they were initially expected to be valid. This creates suspicions that proxy-SU(3) is used as a classification scheme which is valid for all kinds of nuclei (deformed, transitional, spherical). In the same way that Elliott \cite{Elliott1,Elliott2,Elliott3} classified all nuclei in the sd shell within an SU(3) description, proxy-SU(3) appears to be classifying all nuclei 
in heavier shells within a proxy-SU(3) description. In other words, proxy-SU(3) is  a classification scheme valid for all kinds of nuclei. The difference between a classification scheme and a dynamical symmetry \cite{IA,IVI,FVI} has to be clarified. A classification scheme is valid throughout a shell, while a dynamical symmetry applies
only to some nuclei within a shell. Within the sd shell, the SU(3) classification is valid everywhere, but the SU(3) dynamical symmetry is manifested only in the deformed nuclei. Similarly, within higher shells, the proxy-SU(3) classification is valid everywhere, but the proxy-SU(3) dynamical symmetry is manifested only in the deformed nuclei. If we consider spectra, for example, the proxy-SU(3) Hamiltonian is expected to provide good results only for deformed nuclei. In corroboration of this argument, one should recall that in the first Elliott paper \cite{Elliott1}, the full sd shell is classified not only in terms of SU(3), but also in terms of O(6). 

The fact that proxy-SU(3) predictions \cite{proxy2} appear to be correct outside the region of deformed nuclei suggests that the replacement of orbitals carried out within proxy-SU(3) must have deep roots in the shell model, not affected by deformation. This is the focus of the present work. 

We are going to show that although all 0[110] partner Nilsson orbitals exhibit similar expansions over the spherical shell model basis, 
the particular 0[110] Nilsson orbitals involved in the proxy-SU(3) replacement have an extremely simple structure in the spherical shell model basis, with only one vector of the spherical basis corresponding to each Nilsson state. In this way we are going to show that the 0[110] replacement rule used in the Nilsson asymptotic basis for these specific orbitals is translated into a 
$| \Delta N \Delta l \Delta j \Delta \Omega \rangle =  | 1 1 1 0\rangle$ replacement rule in the spherical shell model basis, where $N$ is the principal quantum number, $l$ and $j$ represent the orbital and the total angular momentum respectively, while $\Omega$ is the projection of the total angular momentim on the $z$-axis.

\section{Expansion of Nilsson orbitals in a spherical basis} 

The Nilsson Hamiltonian, which is based on a harmonic oscillator with cylindrical symmetry to which a spin-orbit term and an angular momentum squared term are added,  reads \cite{Nilsson1,NR}
\begin{equation}
H = H_0 - C {\bf l} \cdot {\bf s} - D ({\bf l}^2- \langle {\bf l}^2 \rangle_N),
\end{equation}
where ${\bf l}$ is the orbital angular momentum, ${\bf s}$ is the spin, $C$ and $D$ are parameters, 
\begin{equation}
\langle {\bf l}^2 \rangle _N = {1\over 2} N(N+3)
\end{equation}
is the average of the square of the orbital angular momentum within the $N$th oscillator shell, 
and 
\begin{equation}
H_0= -{\hbar^2 \over 2 M} \nabla^2 +{M\over 2} (\omega_x^2 x^2 + \omega_y^2 y^2 + \omega_z^2 z^2),
\end{equation}
where  $x$, $y$, $z$ are the coordinates in the system fixed to the nucleus, $\omega_x$, $\omega_y$, $\omega_z$ are the harmonic oscillator angular frequencies along these axes, and $M$ is the nuclear mass. 

$H_0$ is separated into \cite{Nilsson1,NR}
\begin{equation}
H_0= \bar H_0 + H_\epsilon
\end{equation}
where 
\begin{equation}
\bar H_0= {1\over 2} \hbar \omega_0 (-\nabla^2 +r^2), 
\end{equation}
\begin{equation}
H_\epsilon= -\epsilon \hbar \omega_0 {4\over 3} \sqrt{\pi \over 5} r^2 Y_{20}.
\end{equation}
The deformation parameter $\epsilon$ is defined in the case of cylindrical symmetry 
through the equations
\begin{equation}
\omega_x^2 =\omega_y^2 = \omega_0^2 \left(1+{2\over 3} \epsilon \right), \qquad
\omega_z^2 = \omega_0^2 \left(1-{4\over 3} \epsilon \right),
\end{equation}
which is related to the quantity $\beta$ of the collective Bohr Hamiltonian \cite{BM}  by
\begin{equation}
\epsilon \simeq {3\over 2} \sqrt{5 \over 4\pi} \beta \simeq 0.95 \beta . 
\end{equation}
The parameters $C$ and $D$ of Eq. (1) are expressed as 
\begin{equation}
C= 2 \kappa \hbar \omega_0, \qquad D=\kappa \mu \hbar \omega_0.
\end{equation}
In the standard parametrization followed in the literature, the dimensionless parameters $\kappa$ and $\mu$ are used, together with dimensionless oscillator units. 

The eigenfunctions of the central force Hamiltonian $\bar H_0$
can be written as $| N l \Lambda \Sigma \rangle$, where $N$ is the principal quantum number, 
$l$ is the orbital angular momentum, $\Lambda$ is the projection 
of the orbital angular momentum on the $z$ axis, and $\Sigma$ is the projection 
of the spin on the $z$ axis. 

The rest of the Hamiltonian does not commute with ${\bf l}^2$, $l_z$, $s_z$, and it also 
connects states with values of $N$ differing by two. One adopts the approximation of neglecting the 
terms nondiagonal in $N$, since they couple states differing in energy by $2\hbar \omega_0$.
One also remarks that the operator $j_z= l_z + s_z$ does commute with the full Hamiltonian, 
therefore the relevant quantum number, the projection of the total angular momentum on the $z$ axis, 
$\Omega = \Lambda + \Sigma$, can be used for labeling the states. 
Therefore the approximate eigenstates of the full Hamiltonian can be expanded as \cite{deShalit}
\begin{equation}\label{expan}
\chi_{N\Omega} = \sum_{l \Lambda} a_{l \Lambda}^\Omega | N l \Lambda \Sigma \rangle.
\end{equation} 
The coefficients $a_{l \Lambda}^\Omega $ have been calculated for different values 
of the parameter 
\begin{equation}
\eta = {2 \hbar \omega_0 \over C} \epsilon
\end{equation}
and are tabulated in Table I of Ref. \cite{Nilsson1}. 

In Ref. \cite{Nilsson1} the various orbitals are labeled by the integer numbers 1 to 74. 
However, it has become customary to use for the Nilsson orbitals the symbol $\Omega [N n_z \Lambda]$, 
where the asymptotic quantum numbers $N$ (the principal quantum number of the major shell),
$n_z$ (the number of nodes in the wave function in the $z$ direction), $\Lambda$ (the projection 
of the orbital angular momentum on the $z$ axis), and $\Omega$ (the projection 
of the total angular momentum on the $z$ axis) are used. (For even--even nuclei, $\Omega = K$.) 
 The quantum numbers $N$ and $\Lambda$ 
become good quantum numbers only at large deformations, where the spin-orbit interaction term $ {\bf l}\cdot {\bf s}$
and the ${\bf l}^2$ term in the Hamiltonian become negligible \cite{deShalit}. This is why these quantum numbers
are called asymptotic. 

By carefully comparing Fig. 5 of Ref. \cite{Nilsson1} to a modern Nilsson diagram \cite{NR}, 
one can see the correspondence between the integers labeling the orbitals in Ref. \cite{Nilsson1}
and the sets of asymptotic quantum numbers corresponding to each orbital.
In this way one obtains the expansion of Eq. (\ref{expan}) for each Nilsson orbital. 

\section{Wave functions in configuration space} 

The full wave functions corresponding to the vectors $| N l \Lambda \Sigma \rangle$ are 
\begin{equation}\label{Psi}
\Psi_{N l \Lambda \Sigma}= R_{nl} Y_{l \Lambda} f_{s \Sigma},
\end{equation}
where $Y_{l \Lambda}$ are the usual spherical harmonics,  
$f_{s \Sigma}$ are the spinors ${1\choose 0}$ and ${0\choose 1}$ for spin up and spin down respectively,
and $R_{nl}$ is the radial wave function of the 3D harmonic oscillator in spherical coordinates, 
given by \cite{Wolf} (in units in which the mass and the frequency of the oscillator, as well as $\hbar$, are 1)
\begin{equation}
R_{nl}= \sqrt{2 (n!) \over \Gamma \left( n+l +{3\over 2} \right) } e^{-r^2/2} r^l L_n^{l+{1\over 2}}(r^2),
\end{equation}
where $\Gamma(x)$ is the Gamma function, $L_n^l(r^2)$ are the Laguerre polynomials, and $N=2n+l$. 

These wave functions can be connected to these appearing in Eq. (8) of Ref. \cite{Nilsson1} 
through the relation connecting confluent hypergeometric functions and Laguerre polynomials
(p. 149 \cite{Greiner})
\begin{equation}
{}_1F_1(-n; m+1; z) = {n! m!\over (n+m)!} L_n^{(m)}(z),
\end{equation}
leading in the present case to 
\begin{equation}\label{FL}
{}_1F_1\left(-n; l+{3\over 2}; r^2\right) = {n! \left(l+{1\over 2} \right)!\over \left(n+l+{1\over 2}\right)!} L_n^{l+{1\over 2}}(r^2).
\end{equation}

\section{An alternative basis}

The Nilsson vectors used in Eq. (\ref{expan}) can easily be expressed in terms of the total angular 
momentum $j$
\begin{equation}\label{jexp}
|N l \Lambda \Sigma \rangle = \sum_j (l \Lambda {1\over 2} \Sigma | j \Omega) | N l j \Omega\rangle , 
\end{equation}
where a Clebsch--Gordan coefficient appears in the rhs. 

Then the expansion of Eq. (\ref{expan}) can be written as 
\begin{equation}
\chi_{N\Omega}= \sum_{lj} \left[ \sum_\Lambda 
a_{l\Lambda}^\Omega  (l \Lambda {1\over 2} \Sigma | j \Omega)  \right] | N l j \Omega \rangle .
\end{equation}
Actually the sum over $l$ can be omitted. This can be seen as follows. The wave function $\chi_{N\Omega}$ is assumed to have 
a definite parity. As seen from Eq. (\ref{Psi}), the parity of the vectors in the rhs of Eq. (\ref{expan}) is dictated 
by the spherical harmonic. As a consequence, only even or only odd values of $l$ can occur in the expansion. 
Since $j=l\pm {1\over 2}$, we see that $j$ and the parity of $\chi_{N\Omega}$ fix which of the two possible values of $l$ will contribute. 

The expansion of Eq. (\ref{jexp}) can be easily inverted 
\begin{equation}
|N l j \Omega \rangle = \sum_\Sigma (l \Lambda {1\over 2} \Sigma | j \Omega) |N l \Lambda \Sigma \rangle, 
\end{equation}
where there is no summation over $\Lambda$ since $\Omega = \Lambda + \Sigma$. 

\section{An example} \label{example1}

The results reported below are based on standard Nilsson calculations for $\epsilon=0.22$ carried out in the proton 50-82 shell with parameter values 
$\kappa=0.0637$ and $\mu=0.60 $, and in the neutron 82-126 shell with parameter values $\kappa=0.0637 $ and $\mu=0.42 $. The same $\kappa$ and $\mu$ values are used throughout this work. 

The expansion in the $|N l \Lambda \Sigma \rangle$ basis obtained for the ${3\over 2}[541]$ proton level is 
\begin{eqnarray}\label{541}
\left|{3\over 2}[541]\right\rangle = 
0.0371 \left| 5 1 1 {1\over 2} \right\rangle \nonumber \\
+ 0.2321 \left| 5 3 1 {1\over 2} \right\rangle 
+ 0.1129 \left| 5 3 2 -{1\over 2} \right\rangle \nonumber \\
+ 0.8070 \left| 5 5 1 {1\over 2} \right\rangle 
+0.5298  \left| 5 5 2 -{1\over 2} \right\rangle.
\end{eqnarray} 

The expansion in the $|N l \Lambda \Sigma \rangle$ basis obtained for the 0[110] partner of this level, the ${3\over 2}[651]$ neutron level, is 
\begin{eqnarray}\label{651}
\left|{3\over 2}[651]\right\rangle = 
0.0681 \left| 6 2 1 {1\over 2} \right\rangle 
+ 0.0229 \left| 6 2  2 -{1\over 2} \right\rangle \nonumber \\
+ 0.2806 \left| 6 4 1 {1\over 2} \right\rangle
0.1644 \left| 6 4 2 -{1\over 2} \right\rangle \nonumber \\
+ 0.7981 \left| 6 6  1 {1\over 2} \right\rangle 
+0.5458  \left| 6 6  2 -{1\over 2} \right\rangle.
\end{eqnarray} 
 One should notice that 0[110] partnership in the usual Nilsson notation $\Omega [ N n_z \Lambda]$ is translated onto [1100] partnership in the $|N l \Lambda \Sigma \rangle$ basis. 
Then the  similarity between the two expansions is clear.
Terms which are [1100] partners have very similar coefficients. The second term in the rhs of Eq. (\ref{651}), which has no counterpart in Eq. (\ref{541}), has the smallest coefficient within Eq. (\ref{651}). 

The similarity becomes more striking by looking at the expansions of the same orbitals in the 
$|N l j \Omega\rangle$ basis, in which one obtains 

 \begin{eqnarray}\label{541b}
\left|{3\over 2}[541]\right\rangle = 
0.0371 \left| 5 1 {3\over 2} {3\over 2} \right\rangle \nonumber \\
- 0.0286 \left| 5 3  {5\over 2} {3\over 2} \right\rangle  
+0.2565 \left| 5 3  {7\over 2}  {3\over 2} \right\rangle \nonumber \\
- 0.0640 \left| 5 5 {9\over 2} {3\over 2} \right\rangle 
+{\bf 0.9633}  \left| 5 5 {11\over 2} {3\over 2} \right\rangle, 
\end{eqnarray} 

 \begin{eqnarray}\label{651b}
\left|{3\over 2}[651]\right\rangle = 
-0.0100 \left| 6 2  {3\over 2} {3\over 2} \right\rangle 
+0.071  \left| 6 2  {5\over 2} {3\over 2} \right\rangle \nonumber \\
-0.0278 \left| 6 4  {7\over 2}  {3\over 2} \right\rangle 
+ 0.3240 \left| 6 4  {9\over 2} {3\over 2} \right\rangle \nonumber \\
-0.0469  \left| 6 6  {11\over 2} {3\over 2} \right\rangle 
+{\bf 0.9418}  \left| 6 6  {13\over 2} {3\over 2} \right\rangle.
\end{eqnarray} 

 One should notice that 0[110] partnership in the usual Nilsson notation $\Omega [ N n_z \Lambda]$ is translated onto $|1110\rangle$ partnership in the $|N l j \Omega \rangle$ basis.
Then the  similarity between the two expansions is clear.
Terms which are $|1110\rangle$ partners have very similar coefficients. The first term in the rhs of Eq. (\ref{651b}), which has no counterpart in Eq. (\ref{541b}), has the smallest coefficient within Eq. (\ref{651b}). Most importantly, the last term in each expansion is clearly the dominant one. 

It is interesting to check if this feature is valid at other deformation values. In Table 1 the coefficients for two additional deformation values,  0.05 and 0.30, are reported. It is clear that the dominance of the last term in each expansion is independent of the deformation.

\section{Further examples} \label{examples} 

We consider the expansions of all orbitals of the 50-82 proton shell and the 82-126 neutron shell in the 
$|N l j \Omega\rangle$ basis for $\epsilon=0.22$. In Table 2  we list the pairs of orbitals for which dominant terms with coefficients higher than 0.9 are appearing for both of them. The dominant terms with their coefficients are also shown. We see that all pairs in the upper part of the table (above the full line) are formed by the orbitals with highest $j$, which are the ones pushed by the spin-orbit interaction down to the shell below.

In other words, on one hand we are unlucky because the spin-orbit interaction breaks the SU(3) symmetry by pushing certain sets of orbitals from one harmonic oscillator shell into another. However, on the other hand we are lucky, because the sets of orbitals crossing from one harmonic oscillator shell into another
are the ones which possess the simplest possible structure in the shell model basis, corresponding to a very good approximation to a single eigenvector of the shell model basis. Furthermore, the orbitals deserting a given shell by going into the shell below and the intruder orbitals invading the same shell coming from the shell above are $|1110\rangle$ partners. In addition, this simplification occurs for all deformations, small and large. 

In different words, the spin-orbit interaction breaks the SU(3) symmetry of the harmonic oscillator by pushing in each shell the orbitals bearing the highest eigenvalue of the total angular momentum $j$ down to the shell below. By the same token, the shell under discussion is invaded by a bunch of intruder orbitals coming from the shell above. In the Nilsson notation we know that these two border-crossing sets of orbitals are 0[110] partners, bearing the same eigenvalues of projections of orbital angular momentum, spin, and total angular momentum and differing only by one quantum in the $z$-direction and therefore by one in the total number of quanta $N$. It so happens, that these pairs of orbitals possess a very simple structure in the shell model basis, corresponding to a single eigenvector each. Therefore they turn out to be $|1110\rangle$ partners in the shell model basis.     

In relation to the proxy-SU(3) approximation, the above imply that when substituting in a given shell the intruder orbitals by the orbitals which had deserted it, the substitution has a very simple and clear form in the shell model basis, since each orbital (but the one with the highest value of $\Omega$, which is not substituted) is substituted by its Nilsson 0[110] partner, which happens to be its $|1110\rangle$ shell model partner. In other words, the proxy substitution retains a very simple form also in the shell model basis. This happens because the substituted orbitals are the ones with the highest value of the total angular momentum. Nilsson orbitals which are 0[110] partners but do not possess the highest total angular momentum still have similar structure in the shell model basis, but if a substitution were needed, it would have been a more complicated process, since several shell model orbitals with different coefficients would have to be replaced, while in the proxy case only a single shell model orbital with unity coefficient is  replaced by another single shell model orbital with unity coefficient.  

\section{A counterexample} \label{example2}

A counterexample is given in Table 3. The two Nilsson orbitals given in this table are 0[110] partners, but they do not belong to the set carrying the maximum $j$ in the relevant shell. We see that at small deformations there is still a dominant shell model eigenvector in each case, with a coefficient higher than 0.9, but as deformation is increased this dominance is lost. The coefficients do change in a ``coherent'' way with increasing deformation, resulting in quite high overlaps of the two Nilsson orbitals, as pointed out in Ref. \cite{Karampagia}, but each Nilsson orbital at large deformation has non-negligible contributions from several shell model eigenvectors. 

\section{Comparison between cylindrical and spherical bases} \label{compar}

In Table 4 the Nilsson orbitals $\Omega[N n_z \Lambda]$ are listed for each harmonic oscillator shell. The quantum number $n_\rho$ in cylindrical coordinates, in which 
\begin{equation}\label{nrho}
N=2 n_\rho +n_z +\Lambda, 
\end{equation}
 is also shown. Next to each Nilsson orbital, the corresponding shell model $ n l ^j_\Omega$ orbital is shown, together with the quantum number $n_r$ in spherical coordinates, in which $n=n_r+1$ and  
  \begin{equation}\label{nr}
N=2 n_r + l.  
\end{equation}

From Table 4 it is clear that in each shell there is a simple one-to-one correspondence between the shell model orbitals bearing the highest $j$ possible within this shell and their Nilsson counterparts. 
These are, for example, the top 5 orbitals in the sdg shell, the top 6 orbitals in the pfh shell, the top 7 orbitals in the sdgi shell. This correspondence is made simple because for these orbitals $n_r=n_\rho=0$, which implies from Eqs. (\ref{nrho}) and (\ref{nr}) that $l=n_z+\Lambda$.  No such simple correspondence exists for the shell model orbitals with $j$ lower than the maximum value.  

From Table 4 it is also clear that the shell model orbitals bearing the highest $j$ possible within this shell are the only ones which are characterized by pure $\Sigma=+1/2$ in the Nilsson asymptotic framework. The rest of the shell model orbitals correspond in the Nilsson asymptotic framework either to pure $\Sigma=-1/2$, or to a mixture of $\Sigma=+1/2$ and $\Sigma=-1/2$. This observation is in accordance to the fact that the proxy-SU(3) substitution leaves all angular momentum projections (projections of orbital angular momentum, total angular momentum, and spin) unaltered.

\section{Discussion}\label{disc}

The main findings of the present work are listed here.

1) The Nilsson 0[110] pairs used in the substitutions occurring in the proxy-SU(3) scheme have an extremely simple structure in the shell model basis $|N l j \Omega \rangle$, with each Nilsson orbital corresponding to a single shell model eigenvector. This simple structure is valid at all deformations, small and large. 

2) The previous point provides a path for testing the accuracy of the proxy-SU(3) approximation. In a shell model calculation in a given shell, one can replace the shell model eigenvectors corresponding to the intruder orbitals (except the one with the highest value of $\Omega$)  by the shell model eigenvectors corresponding to to orbitals which have escaped into the shell below. The replacement will involve orbitals differing by $|1 1 1 0\rangle$ in the shell model basis. 

3) The simple description in the shell model basis reported in 1) occurs only for the orbitals possessing the highest value of total angular momentum $j$ possible within the given shell. For the rest of the orbitals, an approximate one-to-one correspondence between Nilsson orbitals $\Omega [N n_z \Lambda]$ and shell model orbitals $|N l j \Omega \rangle$ is valid at small deformations, but this simple picture breaks down with increasing deformation. At large deformations, several shell model orbitals contribute substantially to each single Nilsson orbital.  

4) The calculation described in 2) is intended to provide a numerical justification for the quality of the proxy-SU(3) approximation. However, as soon as one is persuaded that the approximate SU(3) symmetry indicated by the proxy-SU(3) replacement is valid in the shell model shells, one can directly use the SU(3) symmetry in the shell model shells, without having to worry about the details of the replacements, the only difference remaining being that the approximate SU(3) shell can accommodate one nucleon pair less than the real shell model shell. 

The 0[110] proton-neutron pairs found in Ref. \cite{Cakirli} to correspond to maximum proton-neutron interaction in the rare earth region are the (7/2[523], 7/2[633]) and (7/2[404],7/2[514]) pairs of Table 2 of the present work. 

In the pioneering work of Federman and Pittel \cite{FP1,FP2,FP3} on the onset of deformation, the role of proton-neutron interaction has been pointed out. The proton-neutron pairs responsible for the creation of deformation in the various mass regions are shown in Table 5. In the beginning of each region, the proton-neutron pairs shown at the left part of the table are the ones contributing to the onset of deformation, while further inside each region   the proton-neutron pairs shown at the right part of the table become impontant. One can see that the latter are Nilsson 0[110] pairs, while the former are 0[020] pairs. Further work on understanding the role of 0[020] pairs is called for.

\newpage

\begin{table}
\centering
\caption{Expansions of Nilsson orbitals $\Omega[N n_z \Lambda]$ in the shell model basis $|N l j \Omega \rangle$ for three different values of the deformation $\epsilon$. The Nilsson orbitals shown possess the highest total angular momentum $j$ in their shell. The existence of a leading shell model eigenvector is evident at all deformations. See section \ref{example1} for further discussion. 
}
\begin{tabular}{ r r r r r r  }
\hline\noalign{\smallskip}
${3\over 2}[541]$ & & & & & \\

$|N l j \Omega \rangle$ & 
$\left| 5 1 {3\over 2} {3\over 2} \right\rangle$ & 
$\left| 5 3 {5\over 2} {3\over 2} \right\rangle$ & 
$\left| 5 3 {7\over 2} {3\over 2} \right\rangle$ & 
$\left| 5 5 {9\over 2} {3\over 2} \right\rangle$ & 
$\left| 5 5 {11\over 2} {3\over 2} \right\rangle$ \\
$\epsilon$ & & & & & \\

\noalign{\smallskip}\hline\noalign{\smallskip}
0.05 & 0.0025 & $-0.0015$ & 0.0641 & $-0.0122$ & 0.9979 \\
0.22 & 0.0371 & $-0.0286$ & 0.2565 & $-0.0640$ & 0.9633 \\
0.30 & 0.0601 & $-0.0506$ & 0.3287 & $-0.0922$ & 0.9366 \\

\noalign{\smallskip}\hline
\end{tabular}

\vskip 0.5cm 

\begin{tabular}{ r r r r r r r }
\hline\noalign{\smallskip}
${3\over 2}[651]$ & & & & & & \\

$|N l j \Omega \rangle$ & 
$\left| 6 2 {3\over 2} {3\over 2} \right\rangle$ & 
$\left| 6 2 {5\over 2} {3\over 2} \right\rangle$ & 
$\left| 6 4 {7\over 2} {3\over 2} \right\rangle$ & 
$\left| 6 4 {9\over 2} {3\over 2} \right\rangle$ & 
$\left| 6 6 {11\over 2} {3\over 2} \right\rangle$ &
$\left| 6 6 {13\over 2} {3\over 2} \right\rangle$ \\

$\epsilon$ & & & & & \\

\noalign{\smallskip}\hline\noalign{\smallskip}

0.05 & $-0.0002$ & 0.0046 & $-0.0013$ & 0.0821 & $-0.0086$ & 0.9966  \\
0.22 & $-0.0100$ & 0.0711 & $-0.0278$ & 0.3240 & $-0.0469$ & 0.9418  \\
0.30 & $-0.0207$ & 0.1149 & $-0.0509$ & 0.4091 & $-0.0687$ & 0.9010  \\

\noalign{\smallskip}\hline
\end{tabular}

\end{table}

\begin{table}
\centering
\caption{Nilsson orbitals $\Omega[N n_z \Lambda]$ are listed along with the leading shell model eigenvectors appearing in their expansions in the shell model basis $| N l j \Omega \rangle$, together with the relevant coefficients for $\epsilon=0.22$. See section  \ref{examples} for further discussion.}
\begin{tabular}{ r r r r  }
\hline\noalign{\smallskip}
$\Omega [N n_z \Lambda]$ & $|N l j \Omega\rangle$ &
$\Omega [N n_z \Lambda]$ & $|N l j \Omega\rangle$ \\
\noalign{\smallskip}\hline\noalign{\smallskip}

& & & \\

${1\over 2}[550]$ & 0.9519 $\left| 5 5 {11\over 2} {1\over 2} \right\rangle$ &  
${1\over 2}[660]$ & 0.9270 $\left| 6 6 {13\over 2} {1\over 2} \right\rangle$ \\ 

& & & \\

${3\over 2}[541]$ & 0.9633 $\left| 5 5 {11\over 2} {3\over 2} \right\rangle$ &  
${3\over 2}[651]$ & 0.9418 $\left| 6 6 {13\over 2} {3\over 2} \right\rangle$ \\ 

& & & \\

${5\over 2}[532]$ & 0.9777 $\left| 5 5 {11\over 2} {5\over 2} \right\rangle$ &  
${5\over 2}[642]$ & 0.9610 $\left| 6 6 {13\over 2} {5\over 2} \right\rangle$ \\ 

& & & \\

${7\over 2}[523]$ & 0.9898 $\left| 5 5 {11\over 2} {7\over 2} \right\rangle$ &  
${7\over 2}[633]$ & 0.9783 $\left| 6 6 {13\over 2} {7\over 2} \right\rangle$ \\ 

& & & \\

${9\over 2}[514]$ & 0.9974 $\left| 5 5 {11\over 2} {9\over 2} \right\rangle$ &  
${9\over 2}[624]$ & 0.9911 $\left| 6 6 {13\over 2} {9\over 2} \right\rangle$ \\ 

& & & \\

${11\over 2}[505]$ & 1.0000 $\left| 5 5 {11\over 2} {11\over 2} \right\rangle$ &  
${11\over 2}[615]$ & 0.9983 $\left| 6 6 {13\over 2} {11\over 2} \right\rangle$ \\ 

& & & \\

\hline

& & & \\

${7\over 2}[404]$ & 0.9958 $\left| 4 4 {7\over 2} {7\over 2} \right\rangle$ &  
${7\over 2}[514]$ & 0.9382 $\left| 5 5 {9\over 2} {7\over 2} \right\rangle$ \\ 

\noalign{\smallskip}\hline
\end{tabular}
\end{table}

\begin{table}
\centering
\caption{Expansions of Nilsson orbitals $\Omega[N n_z \Lambda]$ in the shell model basis $|N l j \Omega \rangle$ for three different values of the deformation $\epsilon$. The Nilsson orbitals shown do not possess the highest total angular momentum $j$ in their shell. The existence of a leading shell model eigenvector is evident at small deformation, but this is not the case any more at higher deformations, at which several shell model eigenvectors make considerable contributions. See section \ref{example2} for further discussion.   
}
\begin{tabular}{ r r r r r r  }
\hline\noalign{\smallskip}
${1\over 2}[431]$ & & & & & \\

$|N l j \Omega \rangle$ & 
$\left| 4 0 {1\over 2} {1\over 2} \right\rangle$ & 
$\left| 4 2 {3\over 2} {1\over 2} \right\rangle$ & 
$\left| 4 2 {5\over 2} {1\over 2} \right\rangle$ & 
$\left| 4 4 {7\over 2} {1\over 2} \right\rangle$ & 
$\left| 4 4 {9\over 2} {1\over 2} \right\rangle$ \\

$\epsilon$ & & & & & \\

\noalign{\smallskip}\hline\noalign{\smallskip}
0.05 & $-0.0213$ & 0.1254 & $-0.0702$ & 0.9893 & 0.0127 \\
0.22 & $-0.2248$ & 0.4393 & $-0.2791$ & 0.8057 & 0.1717 \\
0.30 & $-0.2630$ & 0.5003 & $-0.2458$ & 0.7447 & 0.2559 \\

\noalign{\smallskip}\hline
\end{tabular}

\vskip 0.5cm

\begin{tabular}{ r r r r r r r }
\hline\noalign{\smallskip}
${1\over 2}[541]$ & & & & & & \\

$|N l j \Omega \rangle$ & 
$\left| 5 1 {1\over 2} {1\over 2} \right\rangle$ & 
$\left| 5 1 {3\over 2} {1\over 2} \right\rangle$ & 
$\left| 5 3 {5\over 2} {1\over 2} \right\rangle$ & 
$\left| 5 3 {7\over 2} {1\over 2} \right\rangle$ & 
$\left| 5 5 {9\over 2} {1\over 2} \right\rangle$ &
$\left| 5 5 {11\over 2} {1\over 2} \right\rangle$ \\

$\epsilon$ & & & & & \\

\noalign{\smallskip}\hline\noalign{\smallskip}

0.05 & $-0.0200$ & 0.1770 & $-0.0295$ & 0.9780 & $-0.0446$ & $-0.0944$  \\
0.22 & $-0.2492$ & 0.4619 & $-0.3768$ & 0.5550 & $-0.4161$ & $-0.3185$  \\
0.30 & $-0.3121$ & 0.4331 & $-0.4829$ & 0.3430 & $-0.4789$ & $-0.3671$  \\

\noalign{\smallskip}\hline
\end{tabular}

\end{table}

\begin{table}
\centering
\caption{Correspondence between Nilsson orbitals $\Omega[N n_z \Lambda]$ at large deformation $\epsilon$ and shell model orbitals $n l^j_\Omega$. The quantum numbers $n_\rho$ and $\Sigma=\Omega-\Lambda$ in the cylindrical coordinates used by the Nilsson model and $n_r$ in the spherical coordinates used by the shell model are also shown. See section \ref{compar} for further discussion. Below the name of each shell the orbitals belonging to it appear. }
\begin{tabular}{ r r r r r r r r r r r r r r r   }
\hline\noalign{\smallskip}

$\Omega[N n_z \Lambda]$ & $n_\rho$ & $\Sigma$ & $nl^j_\Omega$ & $n_r$ & 
$\Omega[N n_z \Lambda]$ & $n_\rho$ & $\Sigma$ & $nl^j_\Omega$ & $n_r$ & 
$\Omega[N n_z \Lambda]$ & $n_\rho$ & $\Sigma$ & $nl^j_\Omega$ & $n_r$ \\

\noalign{\smallskip}\hline\noalign{\smallskip}
 sdg  & & & &  & pfh  & & & &  & sdgi & & & & \\ 
1/2[440] &0&+& 1g$^{9/2}$ &0& 1/2[550] &0&+&1h$^{11/2}$ &0& 1/2[660] &0&+&1i$^{13/2}$ & 0\\
3/2[431] &0&+& 1g$^{9/2}$ &0& 3/2[541] &0&+&1h$^{11/2}$ &0& 3/2[651] &0&+&1i$^{13/2}$ & 0\\
5/2[422] &0&+& 1g$^{9/2}$ &0& 5/2[532] &0&+&1h$^{11/2}$ &0& 5/2[642] &0&+&1i$^{13/2}$ & 0\\
7/2[413] &0&+& 1g$^{9/2}$ &0& 7/2[523] &0&+&1h$^{11/2}$ &0& 7/2[633] &0&+&1i$^{13/2}$ & 0\\
9/2[404] &0&+& 1g$^{9/2}$ &0& 9/2[514] &0&+&1h$^{11/2}$ &0& 9/2[624] &0&+&1i$^{13/2}$ & 0\\
         & & &            & &11/2[505] &0&+&1h$^{11/2}$ &0&11/2[615] &0&+&1i$^{13/2}$ & 0\\
1/2[431] &0&$-$& 2d$^{5/2}$ &1&          & & &          & &13/2[606] &0&+&1i$^{13/2}$ & 0\\
3/2[422] &0&$-$& 2d$^{5/2}$ &1& 1/2[541] &0&$-$& 2f$^{7/2}$ &1&          & & &       &  \\
5/2[413] &0&$-$& 2d$^{5/2}$ &1& 3/2[532] &0&$-$& 2f$^{7/2}$ &1& 1/2[651] &0&$-$& 2g$^{9/2}$ &1 \\
7/2[404] &0&$-$& 1g$^{7/2}$ &0& 5/2[523] &0&$-$& 2f$^{7/2}$ &1& 3/2[642] &0&$-$& 2g$^{9/2}$ &1 \\
         & &   &            & & 7/2[514] &0&$-$& 2f$^{7/2}$ &1& 5/2[633] &0&$-$& 2g$^{9/2}$ &1 \\
1/2[420] &1&+& 1g$^{7/2}$ &0& 9/2[505] &0&$-$& 1h$^{9/2}$ &0& 7/2[624] &0&$-$& 2g$^{9/2}$ &1 \\
3/2[411] &1&+& 1g$^{7/2}$ &0&          & &   &            & & 9/2[615] &0&$-$& 2g$^{9/2}$ &1 \\
5/2[402] &1&+& 1g$^{7/2}$ &0& 1/2[530] &1&+& 1h$^{9/2}$ &0&11/2[606] &0&$-$&1i$^{11/2}$ &0 \\
         & & &            & & 3/2[521] &1&+& 1h$^{9/2}$ &0&          & &   &            & \\
1/2[411] &1&$-$& 3s$^{1/2}$ &2& 5/2[512] &1&+& 1h$^{9/2}$ &0& 1/2[640] &1&+&1i$^{11/2}$ &0 \\
3/2[402] &1&$-$& 2d$^{3/2}$ &1& 7/2[503] &1&+& 1h$^{9/2}$ &0& 3/2[631] &1&+&1i$^{11/2}$ &0 \\
          & & &       & &          & & &       & & 5/2[622] &1&+&1i$^{11/2}$ &0 \\ 
1/2[400] &2&+& 2d$^{3/2}$ &1& 1/2[521] &1&$-$& 3p$^{3/2}$ &2& 7/2[613] &1&+&1i$^{11/2}$ &0 \\
          & & &       & & 3/2[512] &1&$-$& 3p$^{3/2}$ &2& 9/2[604] &1&+&1i$^{11/2}$ &0 \\
   s    & & &       & & 5/2[503] &1&$-$& 2f$^{5/2}$ &1&          & & &       & \\
1/2[000] &0&+& 1s$^{1/2}$ &0&           & & &       & & 1/2[631] &1&$-$& 3d$^{5/2}$ &2 \\
          & & &       & & 1/2[510] &2&+& 2f$^{5/2}$ &1& 3/2[622] &1&$-$& 3d$^{5/2}$ &2 \\
 p         & & &       & & 3/2[501] &2&+& 2f$^{5/2}$ &1& 5/2[613] &1&$-$& 3d$^{5/2}$ &2 \\
 1/2[110] &0&+& 1p$^{3/2}$ &0&               & & &       & & 7/2[604] &1&$-$& 2g$^{7/2}$ &1 \\
 3/2[101] &0&+& 1p$^{3/2}$ &0& 1/2[501] &2&$-$& 3p$^{1/2}$ &2&          & & &            & \\
          & & &            & &          & &   &            & & 1/2[620] &2&+& 2g$^{7/2}$ &1 \\
1/2[101] &0&$-$& 1p$^{1/2}$ &0&  pf           & & &       & & 3/2[611] &2&+& 2g$^{7/2}$ &1 \\
         & &  &             & &1/2[330] &0&+& 1f$^{7/2}$ &0               & 5/2[602] &2&+& 2g$^{7/2}$ &1 \\
 sd      & &  &             & &3/2[321] &0&+& 1f$^{7/2}$ &0                &          & & &       & \\
1/2[220] &0& +& 1d$^{5/2}$ &0&5/2[312] &0&+& 1f$^{7/2}$ &0               & 1/2[611] &2&$-$& 4s$^{1/2}$ &3 \\      
3/2[211] &0& +& 1d$^{5/2}$ &0&7/2[303] &0&+& 1f$^{7/2}$ &0  &     3/2[602] &2&$-$& 3d$^{3/2}$ &2 \\
5/2[202] &0& +& 1d$^{5/2}$ &0&                & & &       & &              & &   &            & \\
         & &  &            & &1/2[321] &0&$-$& 2p$^{3/2}$ &1          & 1/2[600] &3&+& 3d$^{3/2}$ &2 \\
1/2[211]&0&$-$& 1d$^{3/2}$ &0&3/2[312] &0&$-$& 2p$^{3/2}$ &1&  & & & & \\
3/2[202]&0&$-$& 1d$^{3/2}$ &0&5/2[303] &0&$-$& 1f$^{5/2}$ &0&   & & & & \\
         & & &             & &         & &   &            & &    & & & & \\ 
1/2[200] &1&+& 2s$^{1/2}$  &1&1/2[310] &1&+& 1f$^{5/2}$ &0& & & & & \\
         & & &             & &3/2[301] &1&+& 1f$^{5/2}$ &0&    & & & & \\
         & & &             & &          & & &       & &         & & & & \\
         & & &             & &1/2[301] &1&$-$& 2p$^{1/2}$ &1&  & & & & \\

\noalign{\smallskip}\hline
\end{tabular}
\end{table}

\begin{table}
\centering
\caption{Pairs of orbitals playing a leading role in the development of deformation in different mass regions of the nuclear chart according to Federman and Pittel \cite{FP1,FP2,FP3}. The pairs on the left part of the table contribute in the beginning of the relevant shell, while the pairs on the right become important further within the shell. See section \ref{disc} for further discussion.}
\begin{tabular}{ r r r r r   }
\hline\noalign{\smallskip}
 &  protons  & neutrons &      protons  &     neutrons  \\ 
\noalign{\smallskip}\hline\noalign{\smallskip}
light       & 1d$^{5/2}$     & 1d$^{3/2}$    &    1d$^{5/2}$      &  1f$^{7/2}$        \\
intermediate& 1g$^{9/2}$     & 1g$^{7/2}$    &    1g$^{9/2}$      &  1h$^{11/2}$       \\
rare earths & 1h$^{11/2}$    & 1h$^{9/2}$    &    1h$^{11/2}$     &  1i$^{13/2}$       \\
actinides   & 1i$^{13/2}$    & 1i$^{11/2}$   &    1i$^{13/2}$     &  1j$^{15/2}$       \\

\noalign{\smallskip}\hline
\end{tabular}
\end{table}

\end{document}